\documentclass[conference]{IEEEtran}

\usepackage{float}
\usepackage{enumitem}
\usepackage{cite}
\usepackage{graphicx,amsmath,amssymb,url,multirow}
\ifCLASSOPTIONcompsoc
  \usepackage[caption=false,font=normalsize,labelfont=sf,textfont=sf]{subfig}
\else
  \usepackage[caption=false,font=footnotesize]{subfig}
\fi

\makeatletter

\begin{document}

\title{Online Security Assessment of Low-Inertia Power Systems: A Real-Time Frequency Stability Tool for the Australian South-West Interconnected System}

\author{\IEEEauthorblockN{Alireza Fereidouni\IEEEauthorrefmark{1}, Julius Susanto\IEEEauthorrefmark{1}, Pierluigi Mancarella\IEEEauthorrefmark{2}, Nicky Hong\IEEEauthorrefmark{1}, Teresa Smit\IEEEauthorrefmark{1}, and Dean Sharafi\IEEEauthorrefmark{1}}
\IEEEauthorblockA{\IEEEauthorrefmark{1}Department of Power System \& Market Planning, Australian Energy Market Operator, Perth, WA 6000, Australia}
\IEEEauthorblockA{\IEEEauthorrefmark{2}Department of Electrical and Electronic Engineering, The University of Melbourne, Melbourne, VIC 3010, Australia}
}

\maketitle

\begin{abstract}
In small/medium-sized isolated power networks with low rotational inertia and high penetration of renewables, generation/load contingency events may cause large frequency excursions, potentially leading to cascading failures and even blackouts. Therefore, it is crucial for system operators to be able to monitor the state of the network in real-time and predict the maximum possible frequency deviations at all times. \\
\indent This paper presents a real-time frequency stability (RTFS) tool developed by the Australian Energy Market  Operator (AEMO) and operationalized in the control room for the South West Interconnected System (SWIS) to ensure that the available spinning reserve is sufficient and fast enough to arrest frequency excursions under any conditions, and particularly low-inertia ones. To reduce the computational burden and complexity of the different turbine-governor models, a simple first-order lag function with two adjustable variables has been used for each of the generator. These adjustable parameters, along with other key model parameters such as load damping and inertia, have been calibrated against actual frequency response using high-speed fault recorder data from historical events. As demonstrated in several case studies, the real-time tool has proven to be accurate at predicting the trajectory of system frequency after credible contingencies, thus suggesting that similar implementations could be carried out elsewhere while power systems worldwide progress towards lower inertia.

\end{abstract}

\vspace{1em}
\begin{IEEEkeywords}
Real-time frequency stability tool, low-inertia power systems, online security assessment, frequency stability, power system operation
\end{IEEEkeywords}


\section{Introduction}
\begin{figure}[t]
\centering
\includegraphics[width=2.5in]{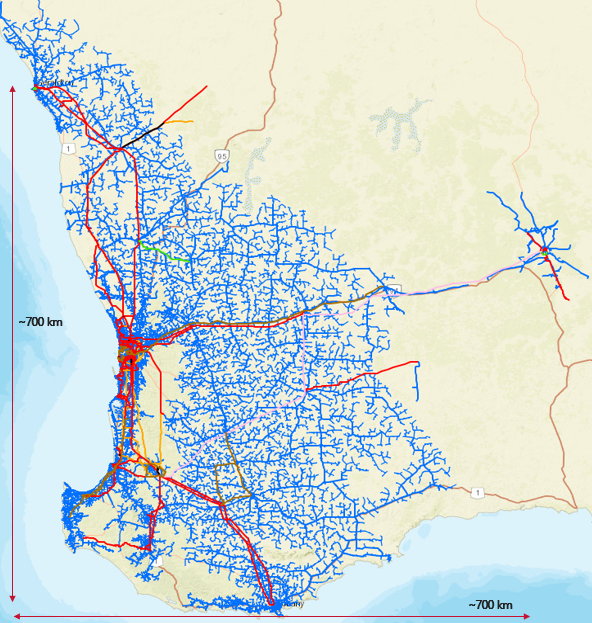}
\caption{Geographic map of the South West Interconnected System (SWIS)}
\label{fig:SWIS}
\end{figure}

\IEEEPARstart{T}{he} need to move towards low-carbon power systems means more and more renewable energy sources are being integrated into electricity grids worldwide. Much  of  this renewable  energy  capacity  is  asynchronously interfaced  to  the  network  via power electronic converters and is thus inertia-less. Large-scale penetration of asynchronous renewables can thus lead to low-inertia conditions and potential frequency instability situations \cite{pierluigi_2017}. This may be particularly severe for small/medium-sized isolated networks. 
The South West Interconnected System (SWIS) is a medium-sized islanded power system serving the southwest region of Western Australia. It is worth pointing out that the SWIS is not connected to the national electricity market (NEM) (i.e., power network in the east coast of Australia). As can be seen from Fig. \ref{fig:SWIS}, the SWIS covers a vast geographic area of 261,000 square kilometres (greater than the entire land area of the United Kingdom), yet the highest peak demand ever recorded in the system is only 4.3 GW (dwarfed by the roughly 60 GW peak demands seen in the National Grid UK system). The SWIS has a rotational  inertia  range  between  8 to  25  GW.s, and it is declining  every  year  due  to continuing  installation  of  renewable  resources  in  the  system. 

\begin{figure}[!htb]
\centering
\includegraphics[width=\linewidth]{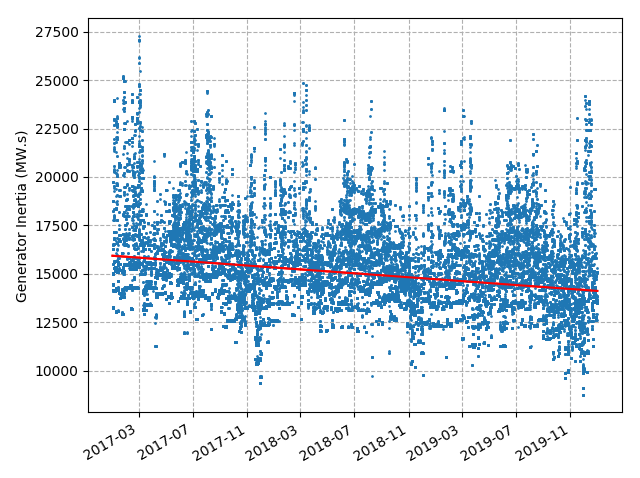}
\caption{Trajectory of synchronous generator inertia in the SWIS from 2017 to 2020}
\label{fig:SWIS_inertia}
\end{figure}

In November 2019, with the recent surge in variable renewable energy generation capacity being added to the system (both small-scale rooftop PV and large-scale wind and solar), the instantaneous penetration of renewable energy in the system rose above 50\% for the first time \cite{aemo_2020}\nocite{Teng_2016}-\cite{Teng:2018:10.1109/ISGTEurope.2017.8260165}. As a result, there has been a noticeable downward trend in synchronous generator inertia in the SWIS over the last few years (as shown in Fig. \ref{fig:SWIS_inertia}).

The quantity of spinning reserves in the SWIS has traditionally been set based on a simple rule of 70\% of the largest generator contingency. This simple provisioning rule has served the system well for many years, but with the downward trend in system inertia, there is a risk that this rule is no longer fit-for-purpose. This is further compounded by the fact that there are generating units in the SWIS with large rated capacities relative to the size of the system. At times, a single generating unit's output can make up nearly 20\% of the system load. Low system inertia coupled with large relative contingency sizes are known to be significant drivers of frequency instability \cite{pierluigi_2018}. Mitigating this risk was the motivation for developing a real-time frequency stability (RTFS) tool that would provide the power system controllers in the SWIS with the situational awareness and insight into the level of frequency instability risk posed at any given time. The main idea of this tool lies in the fact that, as demonstrated in the modelling and studies presented in \cite{pierluigi_2017}, inertia levels and primary frequency response, as well as contingency size \cite{pierluigi_2018}, could be co-optimised so as to guarantee operation within a secure operating envelope for any inertia condition and other relevant parameters such as demand levels, speed of response, etc. 

On the above premises, this  paper  presents  the key  modelling features and implementation aspects  of  the RTFS  tool that has been operationalized in AEMO’s control room of  the SWIS. The tool is based  on  a  low-order  single  mass  machine  model  of  the SWIS. While more sophisticated models may be proposed to study the dynamics of individual generators (see e.g., \cite{Huang_2013}), their system-level deployment for many generators, especially for real-time assessment and control room implementation, is not practical. On the other hand, the use of low-order models to investigate the system frequency response has had a long history in the literature (for example, \cite{Chan_1972} and \cite{anderson_1990}). However the bulk of the earlier studies have been retrospective offline applications where the parameters of the model are calibrated post hoc to align with the measured data, such as in \cite{Inoue_1997}. In \cite{Larsson_2002}, wide-area Phasor Measurement Units (PMU) are used to predict frequency deviations in real-time, but generator primary frequency responses were not included in the model. An inertia monitoring and forecasting tool has also been developed at ERCOT to assess the adequacy of primary frequency response with increasing penetration of renewables \cite{Du_2018}, though there is no relevant work in the literature presenting implementation of a full system dynamic model for online security assessment. 
More recent approaches have applied big data and machine learning techniques to the problem. References \cite{Liu_2018} - \cite{Wang_2019} apply a hybrid model- and data-driven approach to estimate the parameters of a simplified low-order single mass model based on a training data set. Alternatively, authors in \cite{Zhang_2019} - \cite{Li_2020} apply pure machine learning methods to predict the frequency response of the system in a "model-free" manner. However, none of these works demonstrate actual real-world RTFS implementation for online security assessment, especially under challenging low-inertia conditions.

The  key  contribution  of  this  paper  is  the  development  of a mathematical model, parameter validation techniques based on data from real events, and 
demonstration on a system operator’s control room deployment of a system-level tool for RTFS analysis. This online security assessment tool is based on a robust low-order turbine-governor model developed for each droop-enabled machine and tuned empirically as per the real high-speed measurements that can make real-time predictions of  system  frequency  response  in  a  live  operational  setting based  on  available  supervisory  control  and  data  acquisition (SCADA) measurements. The tool also includes online monitoring and forecasting  of the system inertia and estimation of other relevant parameters such as demand-side inertia and load damping factors based on hybrid mathematical and data-driven modelling. The rest of the paper is organized as follows. Section II provides an overview of the proposed frequency stability model, while Section III gives details about the turbine-governor model used for primary frequency response assessment and Section IV describes and exemplifies the methodologies for parameter tuning. Section V then briefly describes the real-time implementation of the model into AEMO’s control room and Section VI demonstrates the very good performance of the tool developed. Finally, Section VII contains the concluding remarks and outlines next steps for tool development.
\begin{figure*}[ht]
\centering
\includegraphics[width=6in]{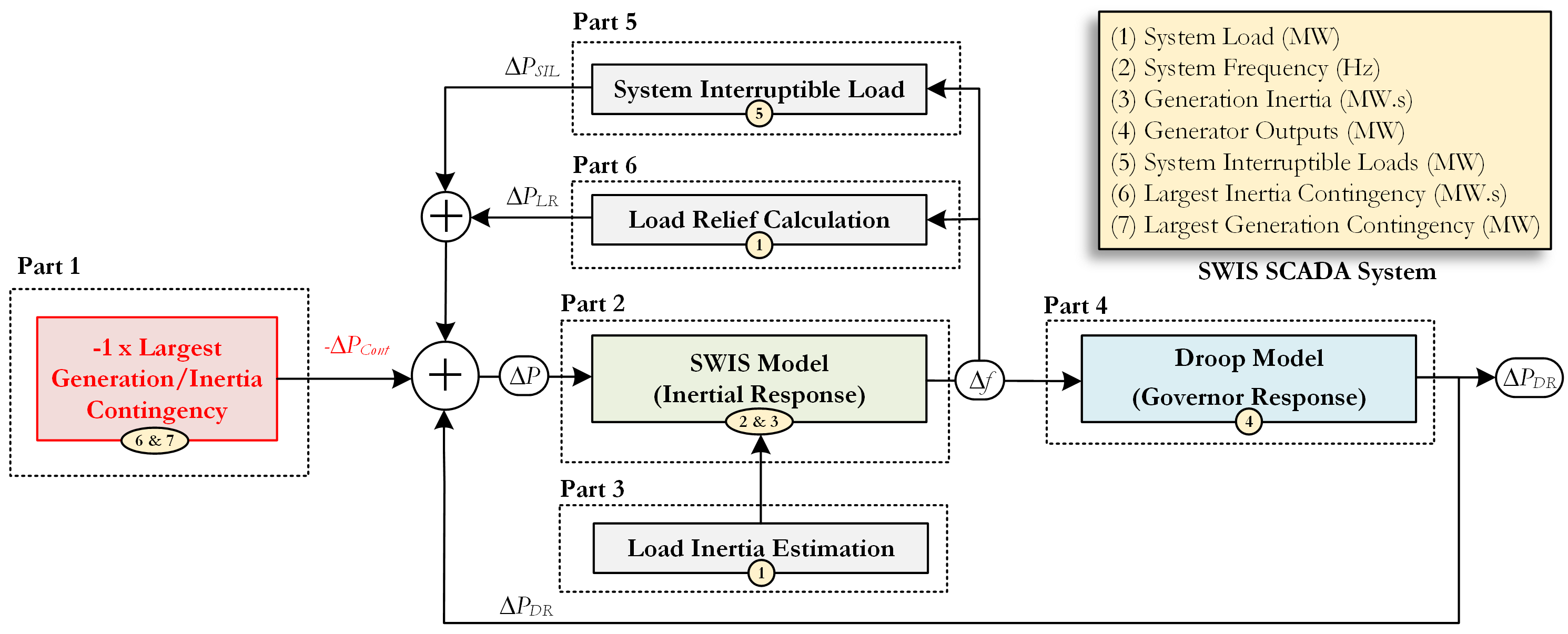}
\caption{Real-time frequency stability (RTFS) model}
\label{fig:RTFS model}
\end{figure*}

\section{Real-Time Frequency Stability Model}
\label{Real-Time Frequency Stability Model}
The developed real-time frequency stability model is shown in Fig. \ref{fig:RTFS model}. This was specifically developed for implementation in the AEMO’s control room of the Australian SWIS to maintain the frequency nadir\footnote{\noindent While the focus on the model description here and current implementation is on low-frequency events and frequency nadir, which are currently more pressing, similar modelling and considerations can be applied to high-frequency events and frequency zenith.} for any single generation contingency events within the acceptable region (i.e., above 48.75 Hz, which is the stage-1 under-frequency load shedding threshold (UFLS) in the SWIS). However, the modelling presented below is relatively general and could easily cater for potential implementation in other systems too. The RTFS model consists of six parts and is explained below.

\subsection{Part 1: Largest Contingency Detection}
From the generator outputs provided by the SCADA system, the model detects the largest online generating unit (in terms of real-time MW output) and largest inertia unit in the SWIS and feeds the MW output of these units ($ \Delta P_{Cont} $) into the model for the calculation of the frequency nadir. The RTFS model toggles the calculation mode between the largest generating unit and largest inertia unit to ensure that the worst-case single credible contingency is evaluated. 

\subsection{Part 2: Inertial Model}
\begin{figure}[t]
\centering
\includegraphics[width=3.5in]{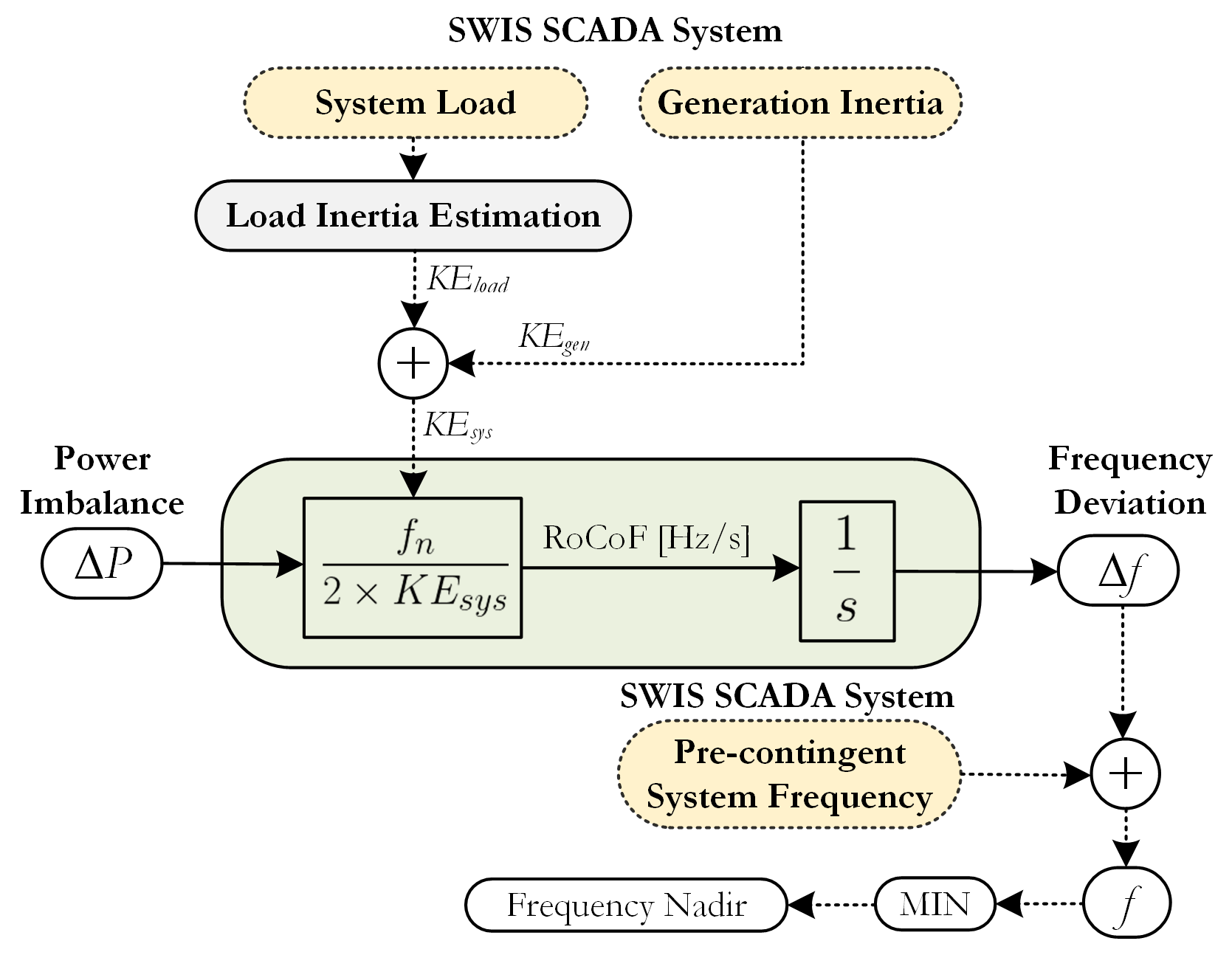}
\caption{RTFS inertial model}
\label{fig:RTFS_inertial model}
\end{figure}

The inertial model, which is constructed based on the single-mass machine theory \cite{anderson_1990}, is shown in Fig. \ref{fig:RTFS_inertial model}. The SWIS is treated as a single machine with a combined kinetic energy of all the online generators under the assumption of uniform frequency throughout the system.

\begin{figure}[ht]
\centering
\subfloat[4 January 2017]{\includegraphics[width=0.48\linewidth]{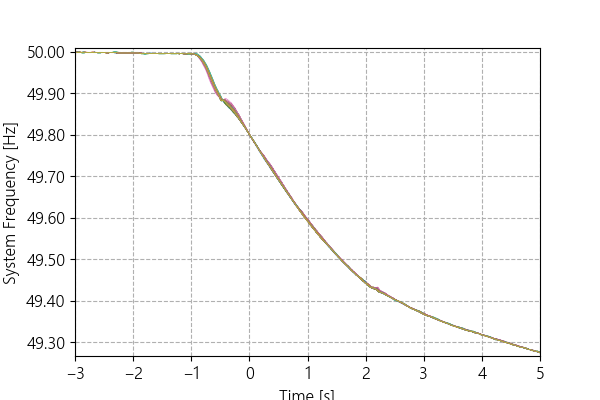}}
\label{fig:COIa}
\subfloat[19 October 2018]{\includegraphics[width=0.48\linewidth]{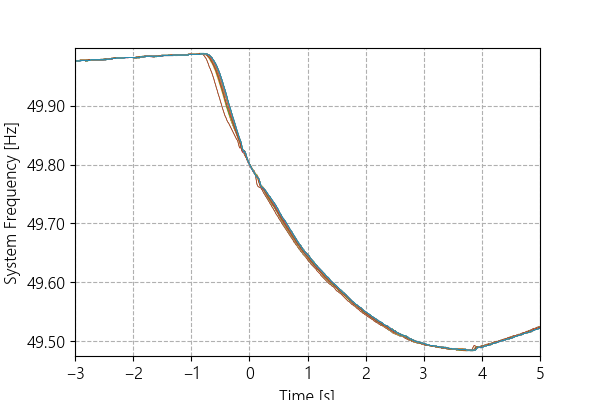}
\label{fig:COIb}}
\hfil
\subfloat[16 March 2019]{\includegraphics[width=0.48\linewidth]{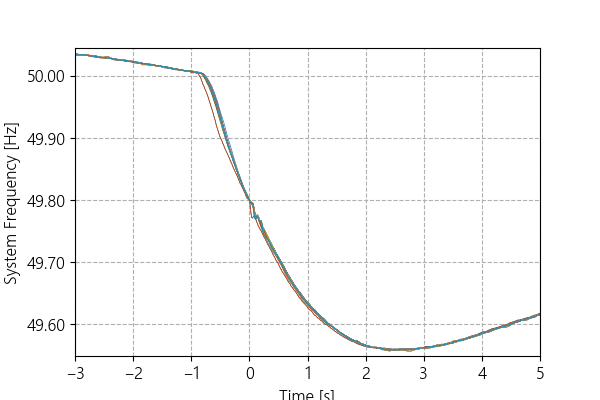}
\label{fig:COIc}}
\subfloat[14 May 2019]{\includegraphics[width=0.48\linewidth]{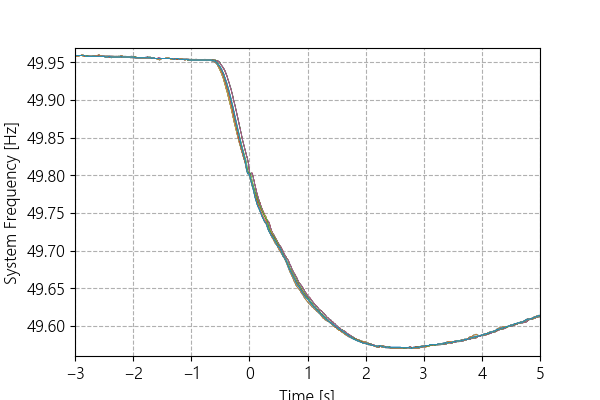}
\label{fig:COId}}
\caption{Examples of SWIS generator frequency trajectories after major disturbances - each figure includes twenty traces measured at different locations}
\label{fig:gen_coherency}
\end{figure}

The use of a low-order aggregate single-frequency model for the SWIS can be justified by examining real system disturbances and the frequencies throughout the system measured using GPS time-synchronised high-speed fault recorders. Fig. \ref{fig:gen_coherency} shows the frequency trajectories of all of the major generators in the SWIS for a sample of four major generator contingencies between 2017 and 2019. It can be seen that the synchronous generators were observed to be broadly coherent system-wide during major disturbances.

The RTFS model uses the swing equation to calculate the frequency deviation as below:

\begin{equation}
\label{eq:swing}
    \begin{split}
\frac{df(t)}{dt} &= \frac{f_n}{2}\times\frac{\Delta P(t)}{KE_{sys}} \\
&= \frac{f_n}{2}\times\frac{\Delta P_{Cont} + \Delta P_{DR}(t) + \Delta P_{LR}(t) + \Delta P_{SDR}(t)}{KE_{load} + KE_{gen}}
    \end{split}
\end{equation}

\noindent where $\frac{df(t)}{dt}$ is the rate of change of frequency in Hz/s (center of inertia), $f_n$ is the SWIS nominal frequency in Hz, $\Delta P(t)$ is the system power imbalance in MW, $KE_{sys}$ is the system total inertia in MW.s, $KE_{load}$ is the total load inertia in MW.s, $KE_{gen}$ is the total generation inertia in MW.s, $\Delta P_{Cont}$ is the contingency size in MW (e.g., -/+300 MW for a generator/load trip), $\Delta P_{DR}(t)$ is the total primary frequency (PFR) response in MW based on the droop response (DR) of each spinning-reserve-provider unit (a positive/negative value for a generator/load trip), $\Delta P_{LR}(t)$ is the load relief in MW (a positive/negative value for a generator/load trip - refer to Section \ref{Part 6: Load Relief Calculation}), and $\Delta P_{SDR}(t)$ is the total amount of automatic system demand response (SDR) in MW (a positive/zero value for a generator/load trip). It is to be noted that $\Delta P_{SDR}(t)$ is procured by AEMO as spinning reserve and is not part of the under-frequency load shedding (UFLS) scheme. The model receives all the information required to calculate the frequency nadir from the SCADA system every four seconds. After solving the differential equation of (\ref{eq:swing}), the frequency nadir is calculated as below:

\begin{equation}
\label{eq:frequency_deviation}
f_{nadir} = \min \bigg(f_0 + \frac{f_n}{2 KE_{sys}}\int \Delta P(t) \ dt \bigg)
\end{equation}

\noindent where $f_0$ is the pre-contingency frequency. The tool calculation engine solves the differential equation given above using the composite trapezoidal method \cite{book_trapz}.
\subsection{Part 3: Load Inertia Estimation}
In the inertial model shown in Fig. \ref{fig:RTFS_inertial model}, the system inertia ($KE_{sys}$) is the sum of total load inertia ($KE_{load}$) and total generation inertia ($KE_{gen}$). $KE_{gen}$ is assumed to be a known parameter calculated, with very good practical results by adding up the nameplate kinetic energy of all the online generators (whose commitment status is available from the SCADA system). But, $KE_{load}$ is not observable from the SCADA system. Therefore, the load inertia is estimated by an empirical equation derived from high speed data recorded for historical generator/load trips - refer to Section \ref{section:load_inertia} for full details.

\subsection{Part 4: Primary Frequency Response Model}
\label{section: part4}
\begin{figure}[ht]
\centering
\includegraphics[width=3.5in]{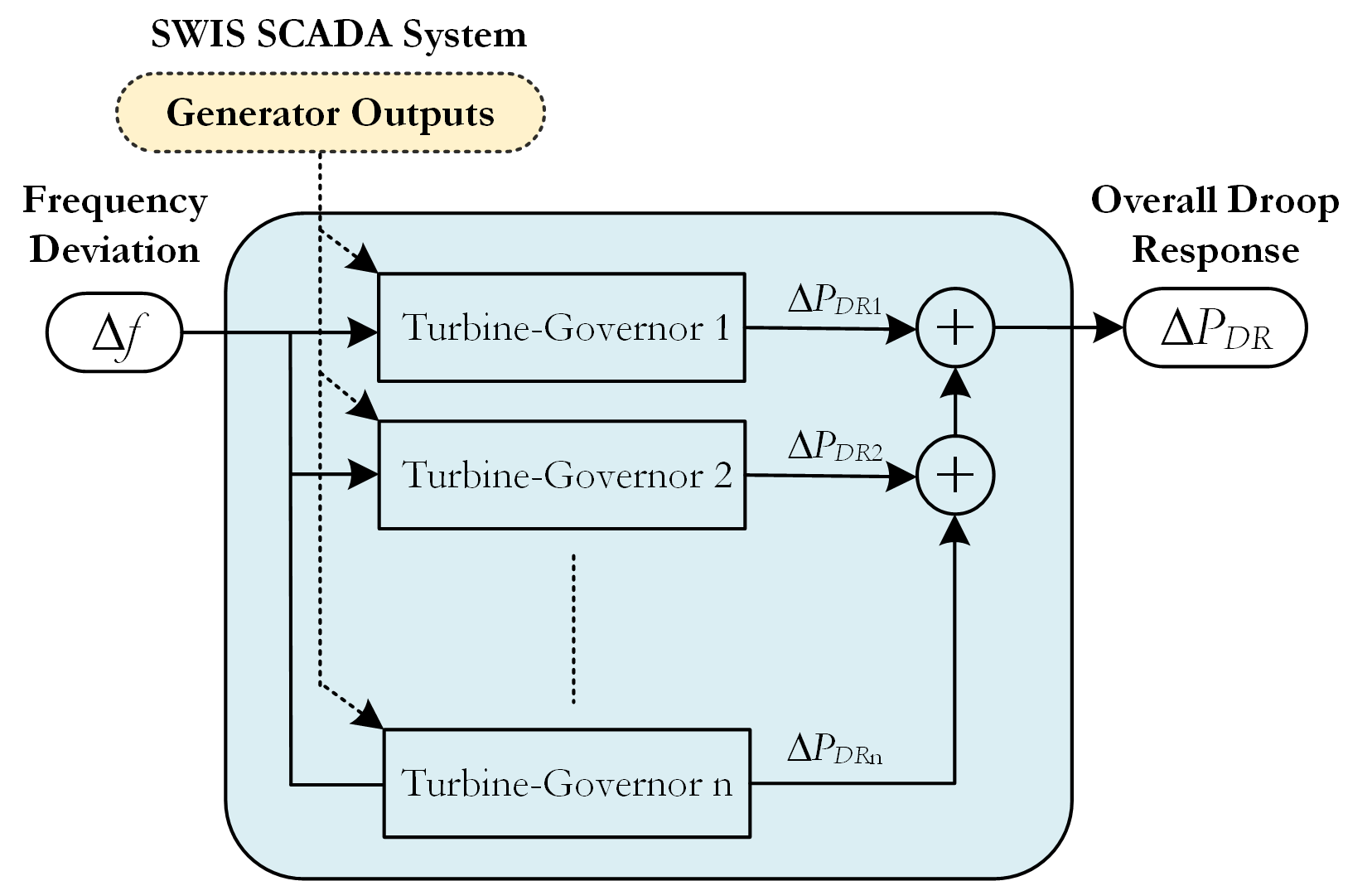}
\caption{RTFS primary frequency response model}
\label{fig:RTFS_PFR}
\end{figure}

In the model shown in Fig. \ref{fig:RTFS_PFR}, $\Delta P_{DR}$ (MW) is the total primary frequency response calculated by aggregating the droop response of all the spinning-reserve-provider units in the SWIS. Fig. \ref{fig:RTFS_PFR} shows the structure of the model used in the RTFS tool - refer to Section \ref{section:PFR} for full details. 

\begin{figure}[t]
\centering
\includegraphics[width=2.5in]{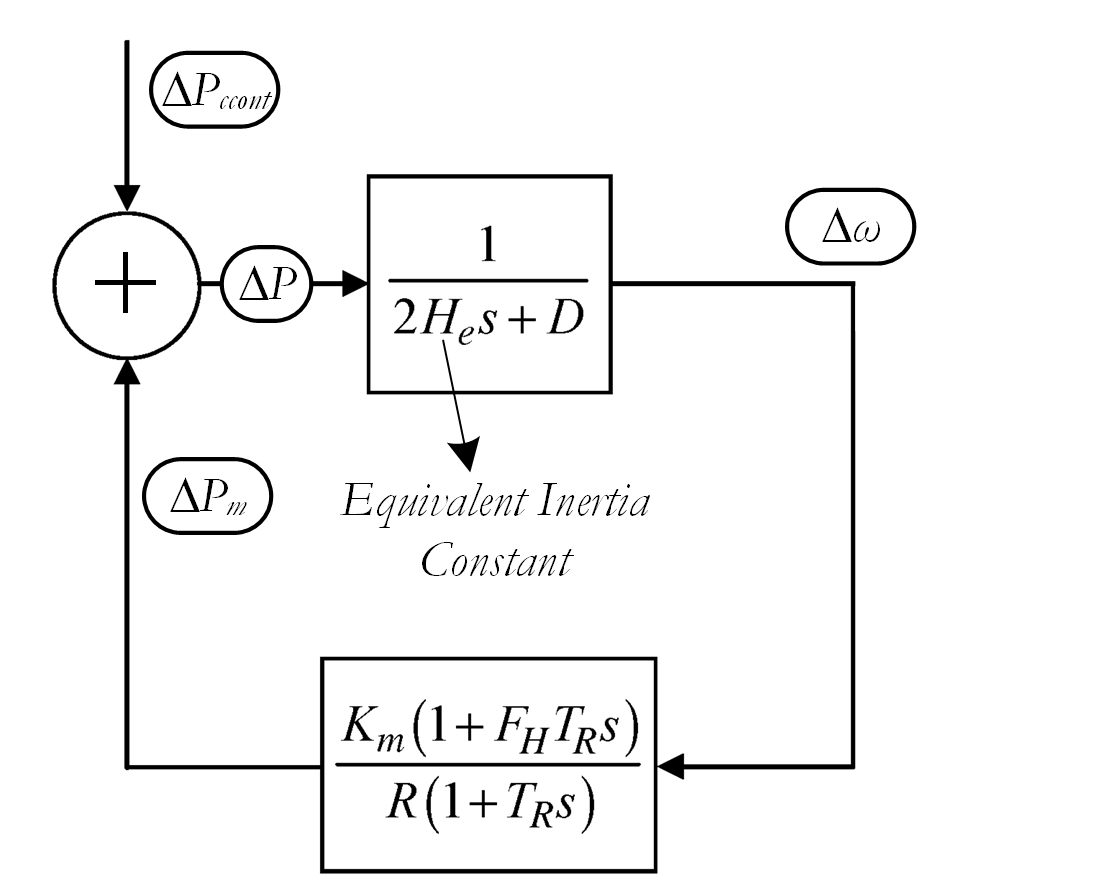}
\caption{Simplified equivalent governor model proposed in \cite{anderson_1990}}
\label{fig:governor_model_reference}
\end{figure}
\subsection{Part 5: Automatic System Demand Response}
Automatic system demand response (SDR) is a contracted predefined amount of load that automatically trips in response to a frequency change in one stage or different stages depending on the under-frequency relay settings and frequency nadir, and it is procured by AEMO as part of spinning reserve (not UFLS). The RTFS model monitors the amount of available SDR in the SWIS and takes it into account for the frequency nadir calculation as $\Delta P_{SDR}$ (MW).

\subsection{Part 6: Load Relief Calculation}
\label{Part 6: Load Relief Calculation}
Load relief value ($\Delta P_{LR}(t)$ in MW) is calculated as per the following equation \cite{susanto_2019}:

\begin{equation}
\label{equation:LRF}
\Delta P_{LR} (t) = P_{load0}\bigg(k_p \times \frac{-\Delta f(t)}{f_n} \bigg)
\end{equation}

\noindent where $P_{load0}$ is the system load at nominal frequency in MW, $k_p$ is the static dimensionless frequency-dependant load relief factor (LRF), $\Delta f(t)$ is the frequency deviation from the nominal frequency in Hz, and $f_n$ is the nominal frequency in Hz.
Given ordered pairs of frequency and system load samples measured from the onset of the contingency to the time of frequency recovery and stabilisation, the load relief factor can be readily estimated by applying a linear regression on (\ref{equation:LRF}) \cite{susanto_2019}.

\section{Turbine-Governor Model for Primary Frequency Response}
\label{section:PFR}
\subsection{Conventional Turbine-Governor Models}
Dynamic models to satisfactorily signify the behaviour of governors, turbines and prime movers have been broadly studied in the literature. Turbine-governor models of steam, hydraulic, gas, and combined cycle turbines are presented in \cite{Mello_1991}\nocite{Mello_1991_coal}–\cite{Inoue_2000}, \cite{Hydraulic_governor}, \cite{Rowen_1983}, and \cite{Mello_1994_gas}, respectively. These models are rather complex, which makes it difficult to execute simple calculations required for analysing the frequency behaviour following generation contingencies.

To reduce the computation burden and complexity of the models, a simplified turbine-governor model has been proposed in \cite{anderson_1990}, and the model is shown in Fig. \ref{fig:governor_model_reference}. This model has proven to be accurate in frequency deviation prediction compared to the detailed models mentioned earlier considering the time scale of the study; However, this model is limited to the reheat steam turbines and all the turbine-governor models must be identical meaning that if the system is equipped with other types of turbines, this model cannot be utilised.
\begin{figure*}[ht]
\centering
\includegraphics[width=6in]{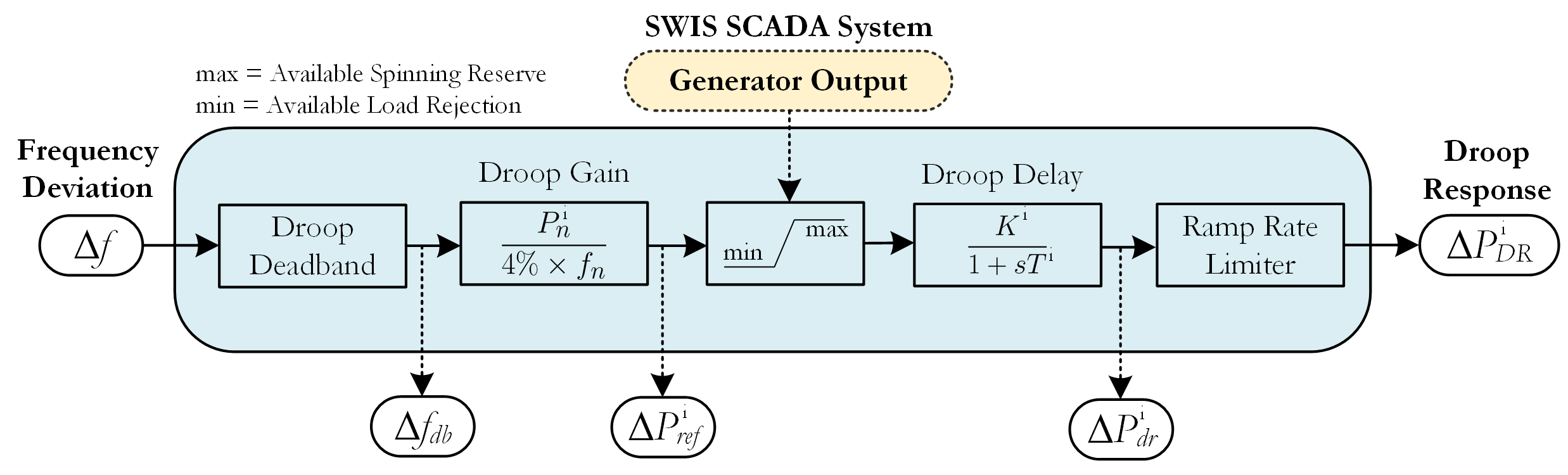}
\caption{Simplified turbine-governor model for each unit ($i$) in the RTFS tool}
\label{fig:RTFS_PFR_Unit}
\end{figure*}

\subsection{Turbine-Governor Model in RTFS Tool}
The turbine-governor model implemented in the tool for each unit in Fig.\ref{fig:RTFS_PFR} is shown in Fig. \ref{fig:RTFS_PFR_Unit}. The proposed model consists of:
\begin{enumerate}[label=\roman*.]

\item \textit{Droop Deadband}: The deadband of a generating unit for droop action in the SWIS must be less than 50 (mHz):
    
\begin{equation}
\label{eq:deadband}
\Delta f_{db}(t) = 
\begin{cases}
\Delta f(t) - 0.025 & \text{if generation contingency}\\
\Delta f(t) + 0.025 & \text{if load contingency}
\end{cases}
\end{equation}

\noindent where $\Delta f_{db}$(t) is the adjusted frequency deviation value after the deadband in Hz and $\Delta f(t) = f_n - f(t)$ in Hz.\\

\item \textit{Droop Gain}:
    In the SWIS, the overall response of a dispatchable unit for power system frequency excursions must be settable and be capable of achieving an increase in the generating's unit active power output of not less than 5\% for a 0.1 (Hz) reduction in power system frequency (i.e., 4\%). Hence, the following equation is used in the model to calculate the droop reference MW for each unit ($i$):
    
\begin{equation}
\label{eq:droop_reference}
\Delta P^i_{ref}(t) = \frac{P^i_n}{4\% \times f_n} \times \Delta f_{db}(t)
\end{equation}  

\noindent where $\Delta P^i_{ref}$(t) is the droop reference value for the governor action in MW of a unit, $P^i_n$ is the MW rating of a unit, $f_n$ is the nominal frequency in Hz, and $\Delta f_{db}$(t) is the adjusted frequency deviation after the deadband.\\

\item \textit{Response Limiter:}
The reference power signal calculated above must not be greater than the unit available spinning reserve or load rejection values for under-frequency or over-frequency events, respectively. Therefore, the reference is capped to the these values, which are calculated as per the data received from the SCADA system.\\

\item \textit{Turbine-Governor Dynamics}:
A simple first-order lag function with two adjustable variables, i.e., $K$ (gain) and $T$ (delay), has been used in the RTFS tool to represent the dynamics of the turbine-governor model. At each instant, the following first-order differential equation is solved in time-domain to calculate $\Delta P^i_{ref}(t)$ for each unit ($i$):

\begin{equation}
\label{eq:droop_diffrential_equation}
\frac{d}{dt}\Delta P^i_{dr}(t)+ \frac{1}{T^i}\Delta P^i_{dr}(t) = \frac{K^i}{T^i}\Delta P^i_{ref}(t)
\end{equation}

It is worth noting that this equation has to be solved for all the droop-enabled units in the system.\\
    \item \textit{Ramp Rate Limiter}
This block limits the unit rate of change of MW to their maximum droop ramp rate (MDRR). At each instant $k$, for each unit $i$, the following equation is used to limit the droop response:

\begin{equation}
\label{eq:droop_ramp_rate_limiter}
\Delta P^i_{DR}[k] =
\begin{cases}
\Delta P^i_{dr}[k]                 & \text{if } \Delta \leq \text{MDRR$^i$}\\
\Delta P^i_{dr}[k-1] + \text{MDRR$^i$} & \text{if } \Delta >    \text{MDRR$^i$}
\end{cases}
\end{equation}

where $\Delta = \Delta P^i_{dr}[k] - \Delta P^i_{dr}[k-1]$.
\end{enumerate}

\section{Model Calibration and Tuning}
The RTFS model is intended to take in SCADA measurements as parameter inputs, but there are a number of key parameters that are not directly observable from SCADA measurements and require careful calibration and tuning in order for the model to be credible for operational use. 

\subsection{System \& Load Inertia Estimation}
\label{section:load_inertia}
System inertia in the SWIS is estimated from historical generator contingency events (filtered only for sudden generator trips as opposed to ramp downs) based on the following equation:

\begin{equation}
KE_{sys} = 0.5 \times f_{n} \times \Delta P \times \bigg[\max\bigg(\frac{d\overline{f}}{dt}\bigg)\bigg]^{-1}
\end{equation}

\noindent where $KE_{sys}$ is the system inertia (in MW.s), $f_n$ is the nominal frequency, $\Delta P$ is the loss in generator output (in MW) and $\frac{d\overline{f}}{dt}$ is the rolling mean of the rate of change in frequency (in Hz/s). 

The rate of change of frequency is estimated from the maximum value of a sample-by-sample time derivative of the high-resolution frequency trace smoothed out by a 500-ms sliding window. This estimation method is in line with the common approach used by ENTSO-E \cite{ENTSOE_2018}, WECC \cite{chassin_2005} and National Grid \cite{ashton_2015}. It was observed that system inertia estimates based on this approach are only accurate for events with high $\frac{df}{dt}$, where the assumption that the initial frequency decline is retarded only by system inertia is valid. For smaller generator trips with low $\frac{df}{dt}$, the influence of primary frequency response becomes more prominent and invalidates this assumption.

The load inertia can be calculated by subtracting the total generator synchronous inertia (which is known from SCADA measurements of which generators are online and a database of inertia constants for each machine):

\begin{equation}
KE_{load} = KE_{sys} - KE_{gen}
\end{equation}

\noindent where $KE_{load}$ is the load inertia (in MW.s) and $KE_{gen}$ is the total synchronous generator inertia (in MW.s). 

\begin{figure}[t]
\centering
\includegraphics[width=3in]{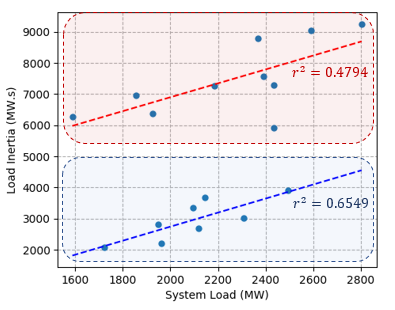}
\caption{Linear regression of historical load inertia estimates vs pre-contingent system load}
\label{fig:load_inertia}
\end{figure}

An estimate for the load inertia was developed for the RTFS model by running a linear regression of historical load inertia estimates against pre-contingent system load. From the scatter plot in Fig. \ref{fig:load_inertia}, it can be seen that while the overall correlations between load inertia and system load are relatively weak ($r^2 = 0.2513$), they tend to fall broadly within two clusters. 

While more data is required to make valid inferences on how load inertia changes with time of day, seasons, etc, it was proposed that a linear regression based on the more conservative lower cluster be used for the model:

\begin{equation}
KE_{load} = 2.2528 \times (P_{load0} - 783)
\end{equation}
\noindent where $P_{load0}$ is the system load at nominal frequency in MW.

\subsection{Load Relief Factor}
An LRF of $k_p = 2$ is used in the model based on a series of estimates made with high-resolution fault recorder data from previous contingency events in the SWIS \cite{susanto_2019} (see also Section. \ref{Part 6: Load Relief Calculation}). This factor has proven to be a fairly robust estimate across a number of events following the operational deployment of the model, though it should be noted that the LRF is continuously being re-estimated to capture trends in load composition and behaviour (e.g. the replacement of directly-coupled induction machines with inverter-fed machines, deindustrialisation, etc). 

\subsection{Generator Primary Frequency Responses}

In the RTFS model, generator governor and turbine / prime mover dynamics are modelled as simplified first-order lag blocks, as from Section. \ref{Real-Time Frequency Stability Model}. While this will not accurately capture all the dynamics of the governor and prime mover under different conditions, it has proven to yield sufficient accuracy for real-time frequency stability studies in the SWIS. 

\begin{figure}[!t]
\centering
\subfloat[Aeroderivative gas turbine unit]{\includegraphics[width=0.48\linewidth]{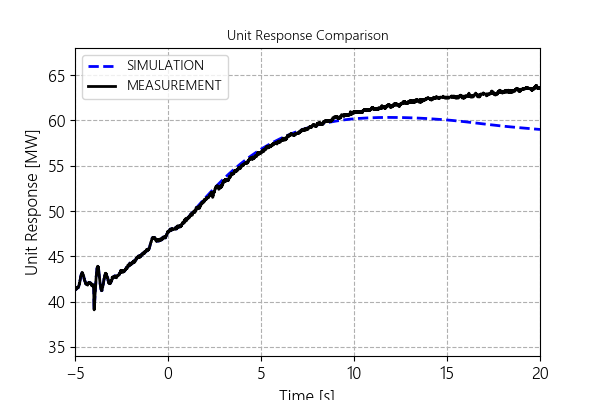}}
\label{fig:calibration_a}
\subfloat[Coal-fired unit]{\includegraphics[width=0.48\linewidth]{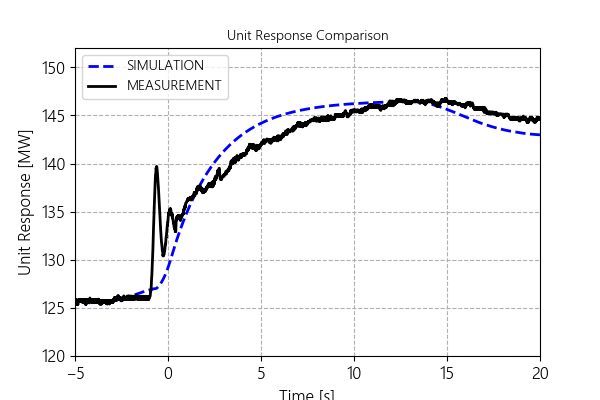}
\label{fig:calibration_b}}
\hfil
\subfloat[Coal-fired unit]{\includegraphics[width=0.48\linewidth]{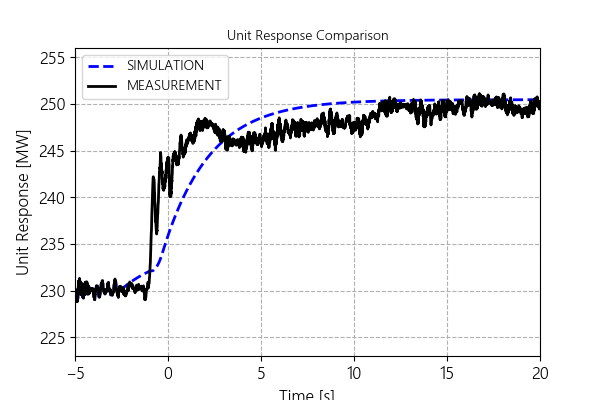}
\label{fig:calibration_c}}
\subfloat[Indutrial gas turnine unit]{\includegraphics[width=0.48\linewidth]{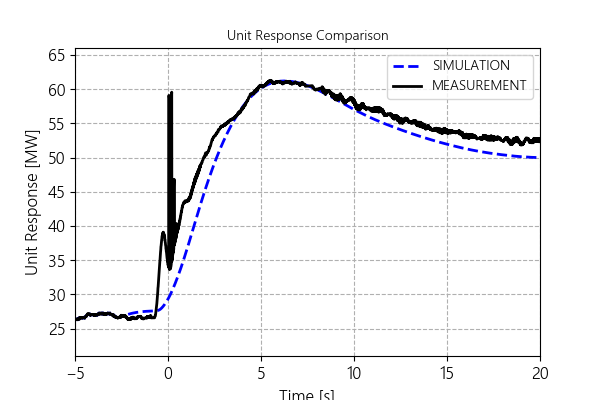}
\label{fig:calibration_d}}
\caption{Calibration of unit primary frequency response parameters}
\label{fig:unit_calibration}
\end{figure}

The simplified models are calibrated using high-speed fault recorder data from historical events, biased towards larger contingency events to understand the potential of each unit under worst-case conditions (i.e. largest possible frequency deviation and rate of change of frequency). Fig. \ref{fig:unit_calibration} shows a sample of the unit responses calibrated against real events. 

It can be seen that in some cases, such as in Fig. \ref{fig:unit_calibration}(a), the actual response deviates from its implied droop target. This is due to supplementary controls such as AGC and/or special local governor control functions. The RTFS model does not capture these supplementary controls and tends to be more conservative as a result. However, overall the model is adequate to capture the main response dynamics in the initial seconds, which is key to maintain the frequency excursion within the desired limits.

\section{Real-time Operationalisation}

The RTFS model was deployed in a live operational environment as a standalone application within the SWIS control room. The application was developed with a graphical user interface along with the following features:

\begin{itemize}
    \item Automatic real-time calculation of system frequency response every 5 minutes
    \item Automatic toggling between the largest generator contingency and largest inertia contingency
    \item Audible and visual alarms if predicted frequency stability limits are breached
    \item A \textit{test mode} that allows the controller to test the frequency response of hypothetical generator redispatch scenarios should an alarm occur
\end{itemize}

Application-specific training was provided to the controllers, as well as operational guidance on how to resolve a frequency stability alarm, e.g. order of precedence for generator re-dispatch.

\section{Performance Evaluation}
In this section, several case studies are presented from recent generator contingency events in the SWIS, evaluating the predictive performance of the RTFS model versus actual measurements.

\subsection{Case Study 1: Gas-fired Power Plant Trip}
In February 2020, there was a trip of a gas-fired power plant in the SWIS resulting in 244 MW of generation loss in the system. Consequently, system frequency declined to a frequency nadir of 49.54 Hz in 2.5 s. Prior to the event, the SWIS was operating in a normal operating state with a total system load and generation inertia of 1.9 GW and 11.5 GW.s, respectively.

\begin{figure}[b]
\centering
\includegraphics[width=3.5in]{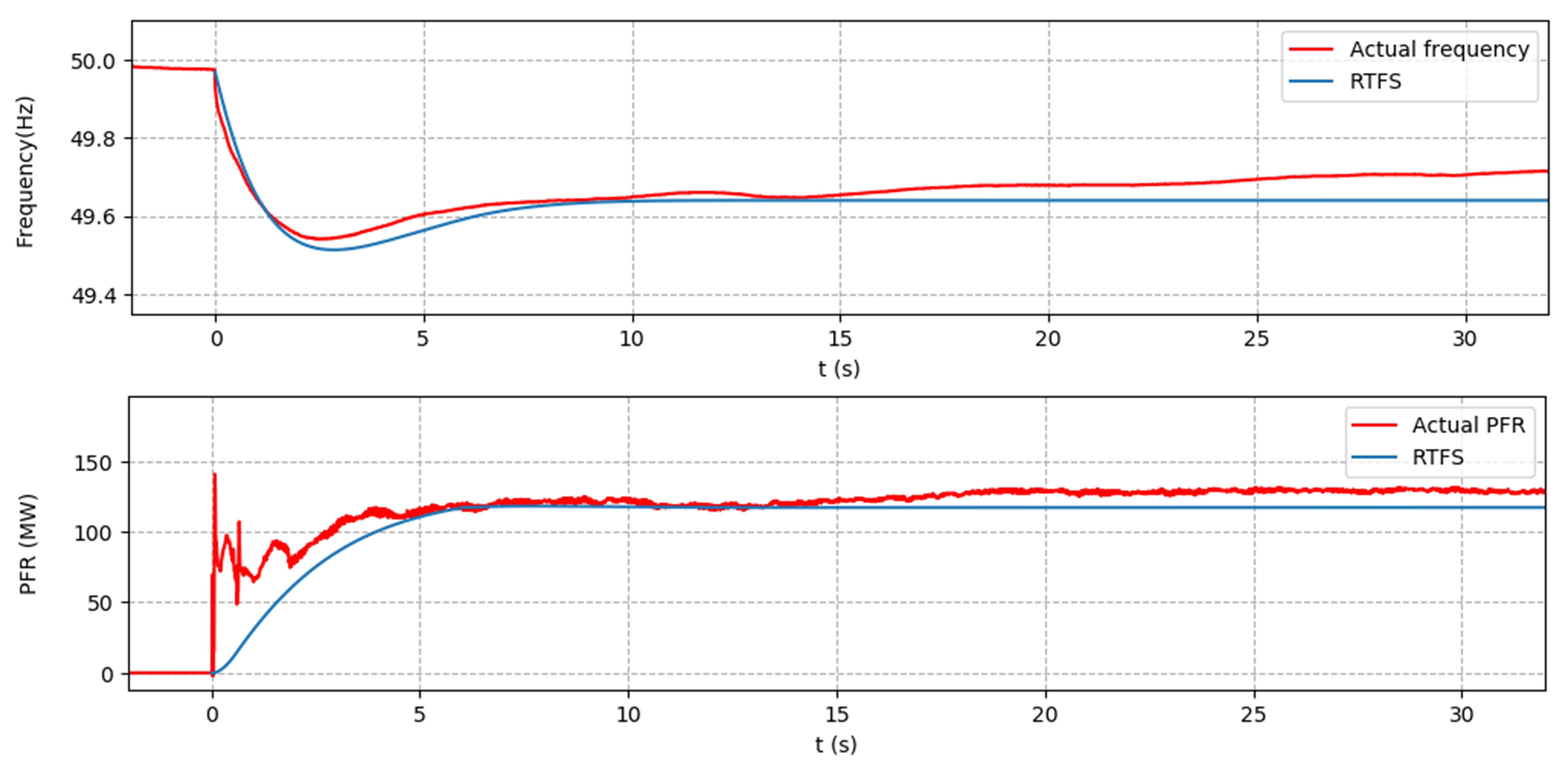}
\caption{Predicted/actual responses for a gas-fired power plant trip}
\label{fig:trip_2020_10_02}
\end{figure}

The frequency traces related to the measurement and RTFS tool (prediction) are shown in Fig. \ref{fig:trip_2020_10_02}. As can be observed from the figure, the RTFS tool, at the time, predicted a frequency nadir of 49.51 Hz, which is quite close to the measurement. It is to be noted that, after approximately 20 s, the curves deviate from each other, and it is due to the response of the units operating on load following to Automatic Generation control (AGC). The overall PFR response (droop) of the units shown in Fig. \ref{fig:trip_2020_10_02} proves the correctness of the RTFS tool calculation as the curves are not only aligned in the amount of MW provided but also in the speed of droop response.

It is worth noting that the spike in the red trace shown in Fig. \ref{fig:trip_2020_10_02} (first 2 seconds) refers to the aggregated inertial response of the generators. which, in the RTFS model, this is captured in the swing equation of (\ref{eq:swing}) using $KE_{sys}$.
\begin{figure}[!htb]
\centering
\subfloat[Original simulation]{\includegraphics[width=0.95\linewidth]{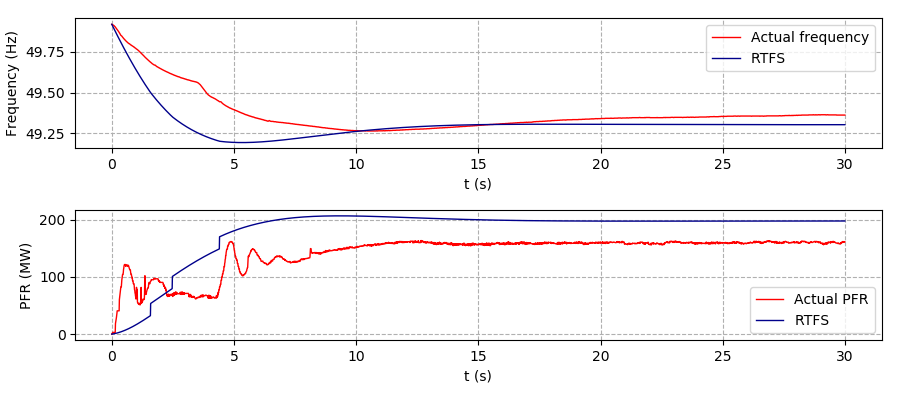}}
\label{fig:ngk_trip_a}
\hfil
\subfloat[Adjusted with steam turbine tripping after gas turbine]{\includegraphics[width=0.95\linewidth]{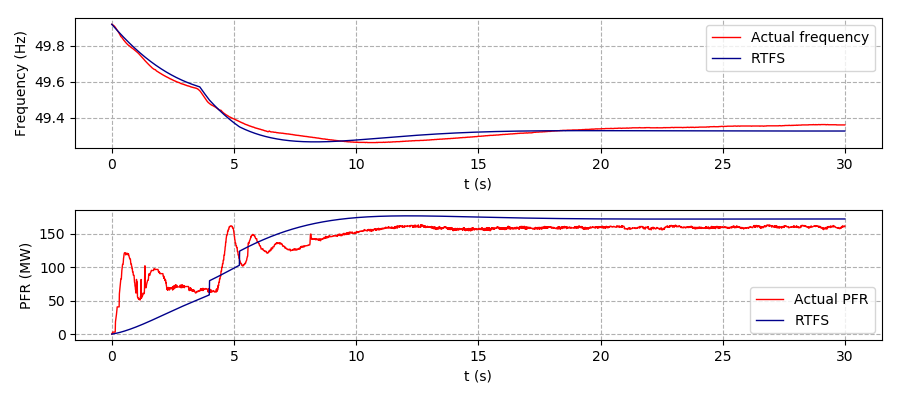}
\label{fig:ngk_trip_b}}
\caption{Predicted/actual responses for a combined-cycle gas power plant trip}
\label{fig:ngk_trip}
\end{figure}
\subsection{Case Study 2: Combined-cycle Gas Power Plant Trip}
This contingency, occurring in January 2020, pertained to the staged disconnection of a 330 MW combined-cycle gas power plant from the SWIS. A mechanical fault had developed in the gas turbine causing it to runback suddenly and then trip. This was followed approximately 4 s later by the trip of the steam turbine. 

Fig. \ref{fig:ngk_trip} shows the actual event traces vs the traces predicted by the RTFS model. The original \textit{out-of-the-box} simulation in Fig. \ref{fig:ngk_trip}(a) is based on a complete instantaneous trip of both the gas and steam turbine units and is conservative as a result, with a frequency nadir that occurs faster and deeper than the actual event. In Fig. \ref{fig:ngk_trip}(b), the simulation was adjusted so that the steam turbine trips after the gas turbine. With these minor adjustments, the resulting predicted frequency and PFR traces are more accurately aligned.

\subsection{Case Study 3: Coal-Fired Power Plant Trip}
\begin{figure}[ht]
\centering
\includegraphics[width=3.5in]{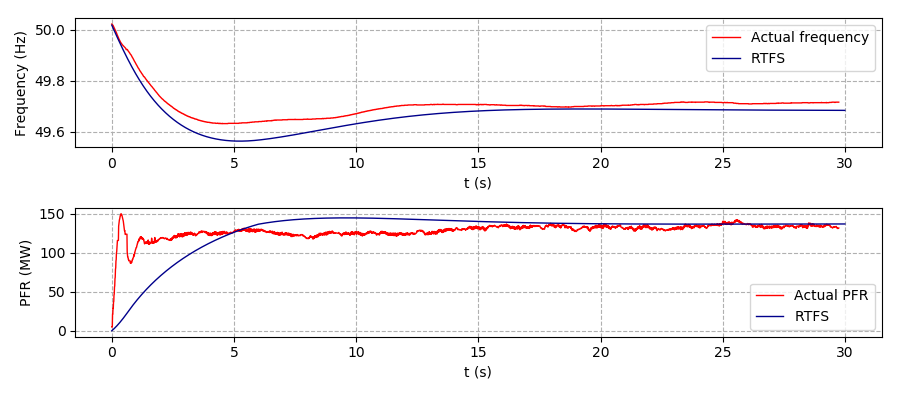}
\caption{Predicted/actual responses for a coal-fired power plant trip}
\label{fig:trip_2019_11_27}
\end{figure}

In this contingency, which occurred in November 2019, a coal-fired power plant tripped resulting in the instantaneous loss of 175 MW. Figure \ref{fig:trip_2019_11_27} shows a comparison of the RTFS tool prediction against actual measurements. 

In this instance, it can be seen that the RTFS tool predicts a more conservative frequency nadir of 49.56 Hz against the actual nadir of 49.63 Hz. This is due to additional droop responses from independent power producers (IPPs) in the SWIS that are required to provide a mandatory droop response to under-frequency events. IPP responses are deliberately excluded from the RTFS tool in order to be conservative as the reliability of these responses is not guaranteed.
\section{Conclusion}
A real-time frequency stability analysis (RTFS) tool, which has been deployed operationally in AEMO's control room of the SWIS for prediction of frequency extrema in real-time, has been presented in this paper. This tool receives all the data required via SCADA system and assists the controllers to better manage the system security by providing information regarding sufficient spinning reserve MW and their speed of response (MW.s) should the worst contingency occurs. To reduce the computational burden and complexity of the different turbine-governor models of the generators in the SWIS, a simple first-order lag function with two adjustable variables has been used for each spinning-reserve provider unit in the tool. Theses adjustable parameters have been calibrated/tuned using high-speed fault recorder data from historical events. As seen from the case studies, the RTFS tool has been successful in accurately predicting the trajectory of system frequency after generator contingencies when compared to actual measurements. This success story suggests that similar implementations could be carried out by system operators in other countries while grid integration of asynchronous renewables drive the system towards lower inertia conditions. Work in progress aims at integrating faster response from emerging resources, e.g., battery storage, into the RTFS tool as well as at coupling the online security assessment to market operation.
 

\bibliographystyle{IEEEtran}
\bibliography{bib_refs}

\end{document}